\documentstyle[preprint,aps,psfig]{revtex}
\newcommand{\beq}{\begin{equation}}
\newcommand{\eeq}{\end{equation}}
\newcommand{\beqa}{\begin{eqnarray}}
\newcommand{\eeqa}{\end{eqnarray}}
\newcommand{\ba}{\begin{array}}
\newcommand{\ea}{\end{array}}

\begin{document}
\draft

\widetext 

\title{Pulsed Quantum Tunneling with Matter Waves} 
\author{Luca Salasnich} 
\address{Istituto Nazionale per la Fisica della Materia, 
Unit\`a di Milano Universit\`a, \\ 
Dipartimento di Fisica, Universit\`a di Milano, \\
Via Celoria 16, 20133 Milano, Italy} 

\maketitle

\begin{abstract} 
In this report we investigate the macroscopic quantum tunneling 
of a Bose condensate falling under gravity and scattering 
on a Gaussian barrier that could model a mirror of 
far-detuned sheet of light. We analyze 
the effect of the inter-atomic interaction and that 
of a transverse confining potential. We show that 
the quantum tunneling can be quasi-periodic and in this way 
one could generate coherent Bose condensed atomic pulses. 
In the second part of the report, 
we discuss an effective 1D time-dependent 
non-polynomial nonlinear Schrodinger equation (NPSE), 
which describes cigar-shaped condensates. 
NPSE is obtained from the 3D Gross-Pitaevskii equation 
by using a variational approach. We find that NPSE 
gives much more accurate results than all other effective 
1D equations recently proposed. 

\end{abstract}

\vskip 0.5cm

\narrowtext

\vskip 0.5cm 

\section{Introduction} 

A macroscopic signature of quantum properties of 
matter is the tunneling of a many-particle Bose condensate 
through a barrier. This subject has been investigated 
by various authors [1-2], in particular in connection to the 
formation of a Josephson current between two wells separated 
by a potential barrier [3-6]. 
\par 
The case we study is that of a falling condensate that 
scatters on a potential barrier that could model 
a mirror formed by a far-detuned sheet of light [7]. 
We find that the interatomic interaction and the 
geometrical aspects of the system, like the aspect ratio 
of the cloud or the fact the the cloud remains trapped 
in the transverse directions, have a strong effect on the 
tunneling probability. 
Moreover, we show that in our system 
macroscopic quantum tunneling (MQT) is a quasi-periodic phenomenon and 
it can be used to generate Bose condensed atomic pulses [8]. 
Finally, from the 3D Gross-Pitaevskii equation and 
using a variational approach, we derive an effective 
1D wave-equation that describes the axial 
dynamics of a Bose condensate confined in an external 
potential with cylindrical 
symmetry. The trapping potential is harmonic in the transverse direction 
and generic in the axial one. Our equation, that is a time-dependent 
non-polynomial nonlinear Schr\"odinger equation (NPSE), 
can be used to model cigar-shaped condensates. 
In the limiting cases of weak and strong interaction, 
our approach gives rise to Schr\"odinger-like equations 
with different polynomial nonlinearities. 

\section{MQT of a Falling Bose Condensate} 

The Bose condensate of a dilute gas 
at zero temperature is well described by the 
3D Gross-Pitaevskii equation (3D GPE) 
\beq
i\hbar {\partial \over \partial t}\psi ({\bf r},t)= 
\left[ -{\hbar^2\over 2m} \nabla^2 
+ V_{ext}({\bf r}) + g (N-1) 
|\psi ({\bf r},t)|^2 \right] \psi({\bf r},t)  \; ,   
\eeq
where $\psi ({\bf r},t)$ is the order parameter 
of the condensate in an external potential 
$V_{ext}({\bf r})$, and $g={4\pi \hbar^2 a_s/m}$, 
with $a_s$ the s-wave scattering length. $N$ is the number 
of condensed atoms. 
\par 
The specific system we consider is a condensate falling under gravity 
and scattering on a Gaussian potential barrier. 
At $t\leq 0$ the condensate is confined by a harmonic trap and 
the external potential reads 
\beq 
V_{ext}(\rho,z)={m\over 2} (\omega_{\rho}^2\rho^2 
+\omega_z^2(z-z_0)^2) + m g z + U e^{-{z^2\over \sigma^2}} \; , 
\eeq
where $\rho=(x^2+y^2)^{1/2}$. 
We use harmonic oscillator units: 
$\omega_H = (\omega_{\rho}^2\omega_z)^{1/3} 
= 2\pi \times 100$ Hz. 
$^{23}$Na atoms with $a_H=(\hbar / (m \omega_H))^{1/2} = 27$ $\mu$m and 
$a_s=3$ nm. 
\par 
We set $z_0=15 a_H$ so that 
the condensate is initially far from the Gaussian 
potential barrier. The 3D GPE is numerically 
solved by using a predictor-corrector splitting method [8]. 
We calculate the ground-state wave-function in the 
external potential by means of the splitting method 
with imaginary time. 
Then, we switch off the harmonic potential 
and use the previous wave-function as initial 
condition. The total energy per particle 
of the condensate is about $180$ $\hbar\omega_H$ with $N$ 
ranging from $1$ to $10^{5}$. 
\par
For large values of the energy barrier 
there is pure bouncing with interference. 
This effect have been experimentally observed by 
Bongs {\it et al} [7]. By reducing the energy barrier, 
in addition to bouncing there is MQT. We have found that,  
if initially the cloud is spherical under free fall, 
the tunneling fraction grows with the number of particles, 
i.e. the chemical potential. Moreover, as shown by 
Fig. 2, MQT is quasi-periodic phenomenon 
due to the quasi-periodic bouncing of a reflected part 
of the condensate, that is also expanding. 
Such a mechanism could be used to generate 
Bose condensed atomic pulses. 
\par 
Fig. 3 shows that the tunneling fraction is reduced 
by the interatomic interaction if the initially spherical 
falling condensate is under a strong enough transverse 
confinement. Instead, for a cigar-shaped quasi-1D 
Bose condensate the interatomic interaction enhances 
the tunneling rate, 
in agreement with the theoretical predictions of Ref. [3]. 

\section{Effective 1D Equation for cigar-shaped condensates}
 
As previously stated, the 3D Gross-Pitaevskii equation (3D GPE) 
is accurate to describe a Bose condensate 
of dilute bosons near zero temperature, where thermal 
excitations can be neglected. 
An interesting problem is to find a reliable mapping from 
the 3D GPE to an effective 1D equation that describes 
cigar-shaped Bose condensates. 
This problem is non trivial due to the nonlinearity 
of the GPE. In this section we illustrate a new 
variational approach which gives very accurate results. 
\par  
First we observe that the 3D GPE can be obtained 
by minimizing the following action functional 
\beq 
S = \int dt d{\bf r} \; 
\psi^*({\bf r},t) \left[ i\hbar {\partial \over \partial t} 
+ {\hbar^2 \over 2 m} \nabla^2 - V_{ext}({\bf r}) 
-{1\over 2}g N |\psi ({\bf r},t)|^2 \right] \psi({\bf r},t)  \; .  
\eeq
Note that we now use $N$ instead of $N-1$. 
We analyze a trapping potential that is generic 
in the axial direction and harmonic in the radial one: 
\beq
V_{ext}({\bf r})={1\over 2}m\omega_{\bot}^2(x^2+y^2) + V(z) \; 
\eeq
We minimize the action functional $S$ 
by choosing the following trial wavefunction [9]:   
\beq 
\psi({\bf r},t) = \phi(x,y,t;\eta(z,t)) \; f(z,t) \; , 
\eeq
where both $\phi$ and $f$ are normalized and $\phi$ is represented 
by a Gaussian [10]: 
\beq
\phi(x,y,t;\eta(z,t)) = { e^{-(x^2+y^2)\over 2 \eta(z,t)^2} 
\over \pi^{1/2} \eta(z,t)} \; . 
\eeq 
The variational functions $\eta(z,t)$ and $f(z,t)$ 
will be determined by minimizing the action functional 
after integration in the $(x,y)$ plane. 
\par 
From the two Euler-Lagrange equations, one eventually obtains 
$$
i\hbar {\partial \over \partial t}f= 
\left[ -{\hbar^2\over 2m} {\partial^2\over \partial z^2} 
+ V + {g N \over 2\pi a_{\bot}^2} {|f|^2\over \sqrt{1+ 2a_sN|f|^2} } 
\right. 
$$
\beq 
+ {\hbar \omega_{\bot}\over 2}  
\left. 
\left( {1\over \sqrt{1+ 2 a_sN|f|^2} } + \sqrt{1+ 2a_sN|f|^2}
\right) \right] f  \; . 
\eeq
This equation is a time-dependent non-polynomial nonlinear 
Schrodinger equation (NPSE) [10]. 
Note that from NPSE in certain limiting cases 
one recovers familiar results. 
In the weakly-interacting limit $a_sN|f|^2 <<1$ 
the previous equation reduces to 
\beq
i\hbar {\partial \over \partial t}f= 
\left[ -{\hbar^2\over 2m} {\partial^2\over \partial z^2} 
+ V + {g N \over 2\pi a_{\bot}^2} |f|^2 \right] f  \; ,  
\eeq
where the constant $\hbar \omega_{\bot}$ has been omitted because 
it does not affect the dynamics. This equation is a 
1D Gross-Pitaevskii equation (1D GPE), whose 
nonlinear coefficient is $g'=g/(2\pi a_{\bot}^2)$. 
This ansatz has been used by various authors, 
for example in Ref. [11]. 
\par
In the strongly-interacting limit $a_sN|f|^2 >> 1$ 
the NPSE becomes 
\beq
i\hbar {\partial \over \partial t}f= 
\left[ -{\hbar^2\over 2m} {\partial^2\over \partial z^2} 
+ V + {3\over 2}{g N^{1/2} 
\over 2\pi a_{\bot}^2 \sqrt{2a_s}} |f| \right] f  \; .   
\eeq
In this limit, and in the stationary case, 
the kinetic term can be neglected (Thomas-Fermi approximation) 
and one finds 
\beq
|f(z)|^2 = {2\over 9} {1\over (\hbar\omega_{\bot})^2 a_s N} 
\left( \mu' - V(z) \right)^2 \; , 
\eeq
where $\mu'$ is the chemical potential, fixed by the 
normalization condition. This 1D Thomas-Fermi density 
profile is quadratic in the term $\mu' -V(z)$. 
The same quadratic depencence is obtained starting 
from the Thomas-Fermi approximation of the 3D stationary GPE, 
i.e. neglecting the spatial derivatives in Eq. (1), 
and then integrating along $x$ and $y$ variables. 
In this way one finds a formula that differs from 
the previous one only for the numerical 
factor which is $1/4$ instead of $2/9$. 
\par 
To test the accuracy of the NPSE, we take 
$V(z)= (1/2) m \omega_z^2 z^2$, and compare the results of 
4 different procedures: 
1) 3D GPE: numerical solution of 3D GPE; 
2) 1D GPE: numerical solution of 1D GPE. 
3) CGPE:  numerical solution of 1D GPE, where 
the nonlinear term is found by imposing that the 
1D wave-function has the same chemical potential 
of the 3D one [12]; 4) NPSE: numerical solution of NPSE. 
\par 
As shown in Fig. 5, CGPE gives better results than 1D GPE 
for large values of scattering length but in any case 
NPSE is superior. Our calculations (for details see [10]) 
show that the ground-state and also the dynamics 
of the condensate are accurately described by NPSE. 

\section*{Conclusions} 

We have studied the dynamics of Bose condensates 
falling under gravity and scattering on a Gaussian barrier 
that models a mirror of light. 
Apart the pure bouncing with interference, 
that can be seen for very large values 
of the energy barrier, we have investigated 
quantum tunneling. In our system quantum tunneling 
is a quasi-periodic phenomenon: 
it can be used to generate Bose-Einstein  condensed atomic 
pulses. In addition, we have found that 
geometrical aspects are important in 
tunneling: 1D results do not always reflect the behavior 
of 3D systems. 
Finally, we have verified that the 1D 
non-polynomial nonlinear Schr\"odinger equation we have obtained 
by using a Gaussian variational ansatz from the 3D Gross-Pitaevskii 
action functional is quite reliable in describing  
cigar-shaped condensates. The agreement with the 
results of the 3D Gross-Pitaevskii equation is very good 
for both ground-state and dynamics of the condensate. 
This equation will be useful for detailed numerical analysis 
of the dynamics of cigar-shaped condensates, particularly 
when the local density may undergo sudden and large variations. 
In conclusion, we observe that Pulsed MQT of matter waves can be 
generated also with other mechanisms. For example, we are 
currently studying the stimulated MQT of matter waves 
produced by a Bose condensate trapped in a potential with 
a periodically modulated energy barrier. 

\section*{References} 

\begin{description}

\item{\ [1]} Smerzi, A., Fantoni, S., Giovannazzi, S. 
and Shenoy, S.R., 1997, Phys. Rev. Lett. {\bf 79}, 4950. 

\item{\ [2]} Milburn, G.J., Corney, J., Wright, E. and 
Walls, D.F., 1997, Phys. Rev. A {\bf 55}, 4318.  

\item{\ [3]} Zapata, I., Sols, F. and Leggett, A.J., 
1998, Phys. Rev. A {\bf 57}, R28. 

\item{\ [4]} Raghavan, S., Smerzi, A., Fantoni, S., 
and Shenoy, S.R., 1999, Phys. Rev. A {\bf 59}, 620. 

\item{\ [5]} Salasnich, L., Parola, A., and Reatto, 
L., 1999, Phys. Rev. A {\bf 60}, 4171.  

\item{\ [6]} Pozzi, B., Salasnich, L., Parola, A., and Reatto, 
L., 2000, Eur. Phys. J. D {\bf 11}, 367.  

\item{\ [7]} Bongs, K., Burger, S., Birkl, G., Sengstock, K., 
Ertmer, W., Rzazewski, K., Sampera, A., and Lewenstein, M., 
Phys. Rev. Lett. 83, 3577 (1999). 

\item{\ [8]} Salasnich, L., Parola, A., and Reatto, L., 
cond-mat/0105092, to be published in Phys. Rev. A {\bf 64}, 
026XX (2001). 

\item{\ [9]} Jackson, A.D., Kavoulakis, G.M., and Pethick, C.J., 
Phys. Rev. A {\bf 58}, 2417 (1998). 

\item{\ [10]} Salasnich, L., Parola, A., and Reatto, L., 
"Effective wave-equations for the dynamics of 
cigar-shaped and disc-shaped Bose condensates", 
submitted for publication. 

\item{\ [11]} M. Olshanii, Phys. Rev. Lett. {\bf 81}, 938 (1998). 

\item{\ [12]} M.L. Chiofalo and M.P. Tosi, Phys. Lett. A 
{\bf 268}, 406 (2000).

\end{description}

\newpage 

\begin{figure}
\centerline{\psfig{file=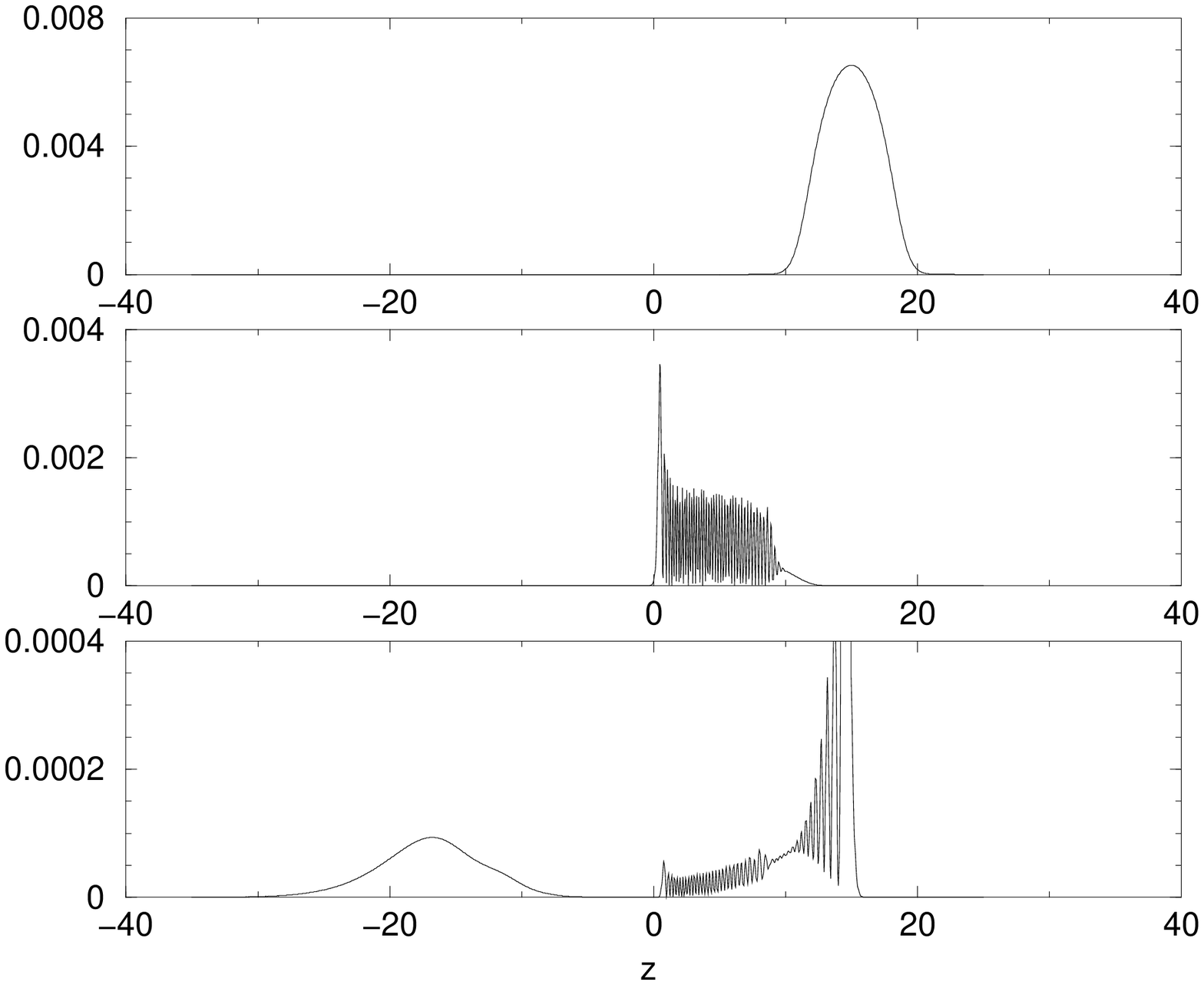,height=5.in}}
{\bf Fig. 1}: {Density profile $|\psi(\rho=0,z)|^2$ 
of the free falling Bose condensate at $t=0$, $t=2$ and $t=4$. 
Gaussian energy barrier $U e^{z^2/\sigma^2}$, 
with $U=200$ and $\sigma=1$. Initial chemical potential:  
$\mu =184.23$. 
$N=10^5$ condensed atoms. 
Lenght in units $a_H=27$ $\mu$m, time in units $\omega_H^{-1}=1.6$ ms and 
energy in units $\hbar \omega_H$.} 
\end{figure} 

\newpage 

\begin{figure}
\centerline{\psfig{file=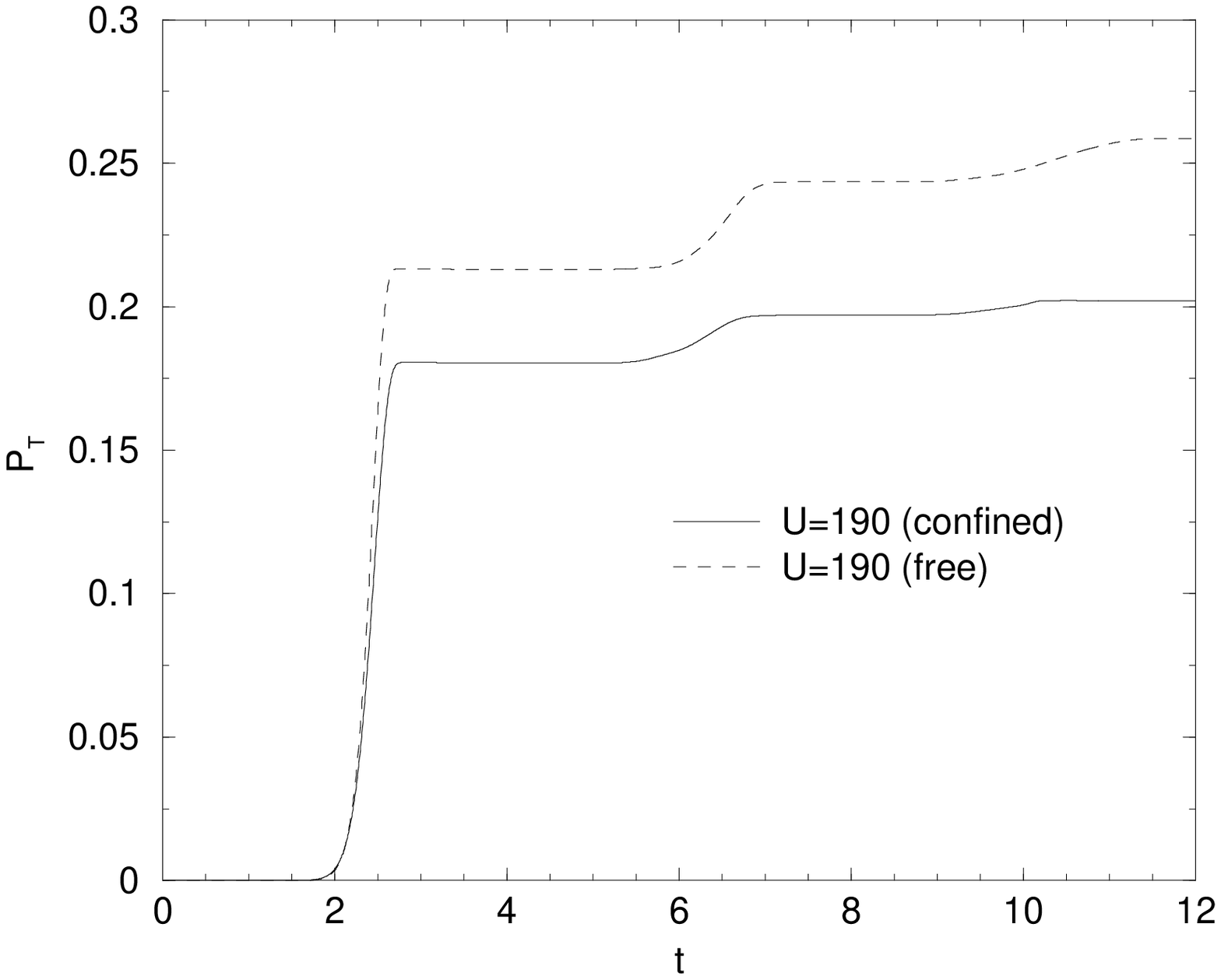,height=5.in}}
{\bf Fig. 2}: {Tunneling fraction $P_T$ as a function of time $t$. 
Comparison between radially-free ($\omega_{\rho}=0$) 
and under radial confinament ($\omega_{\rho}=0.5$) falling 
condensates. $N=10^5$ condensed atoms and 
initial position $z_0=15$.  
Gaussian energy barrier $U e^{z^2/\sigma^2}$, 
with $\sigma =1$. Units as in Fig 1.} 
\end{figure}

\newpage

\begin{figure}
\centerline{\psfig{file=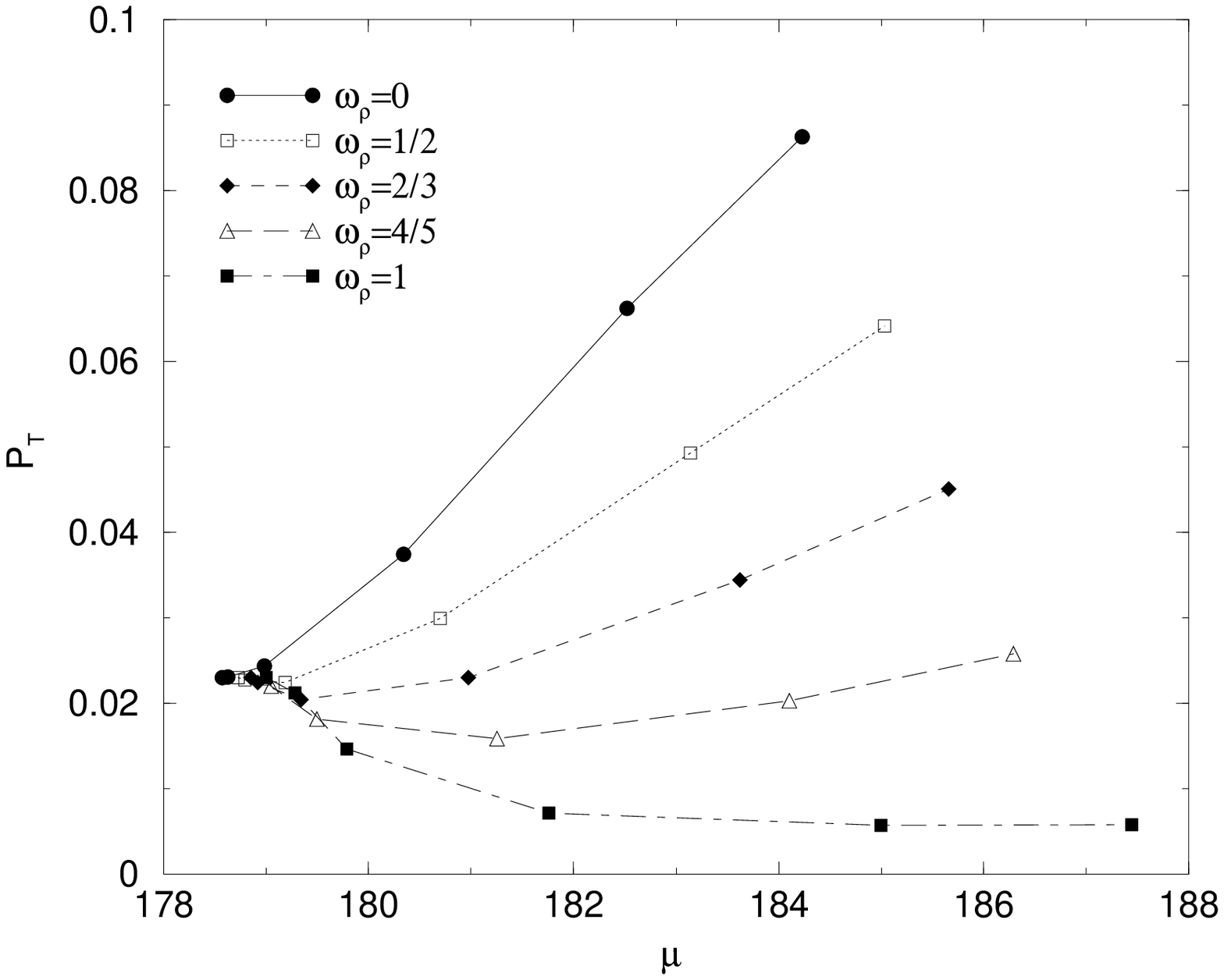,height=5.in}}
{\bf Fig. 3}: {Tunneling fraction $P_T$ as a function of 
the initial chemical potential $\mu$ for different 
frequencies $\omega_{\rho}$ of radial confinament. 
The points in each line correspond to 
a sequence of numbers $N$ of atoms. From left to right: 
$N=1,10^2,10^3,10^4,5\times 10^4,10^5$. 
Gaussian energy barrier $U e^{z^2/\sigma^2}$, 
with $U=200$ and $\sigma=1$. 
Units as in Fig 1.} 
\end{figure} 

\newpage

\begin{figure}
\centerline{\psfig{file=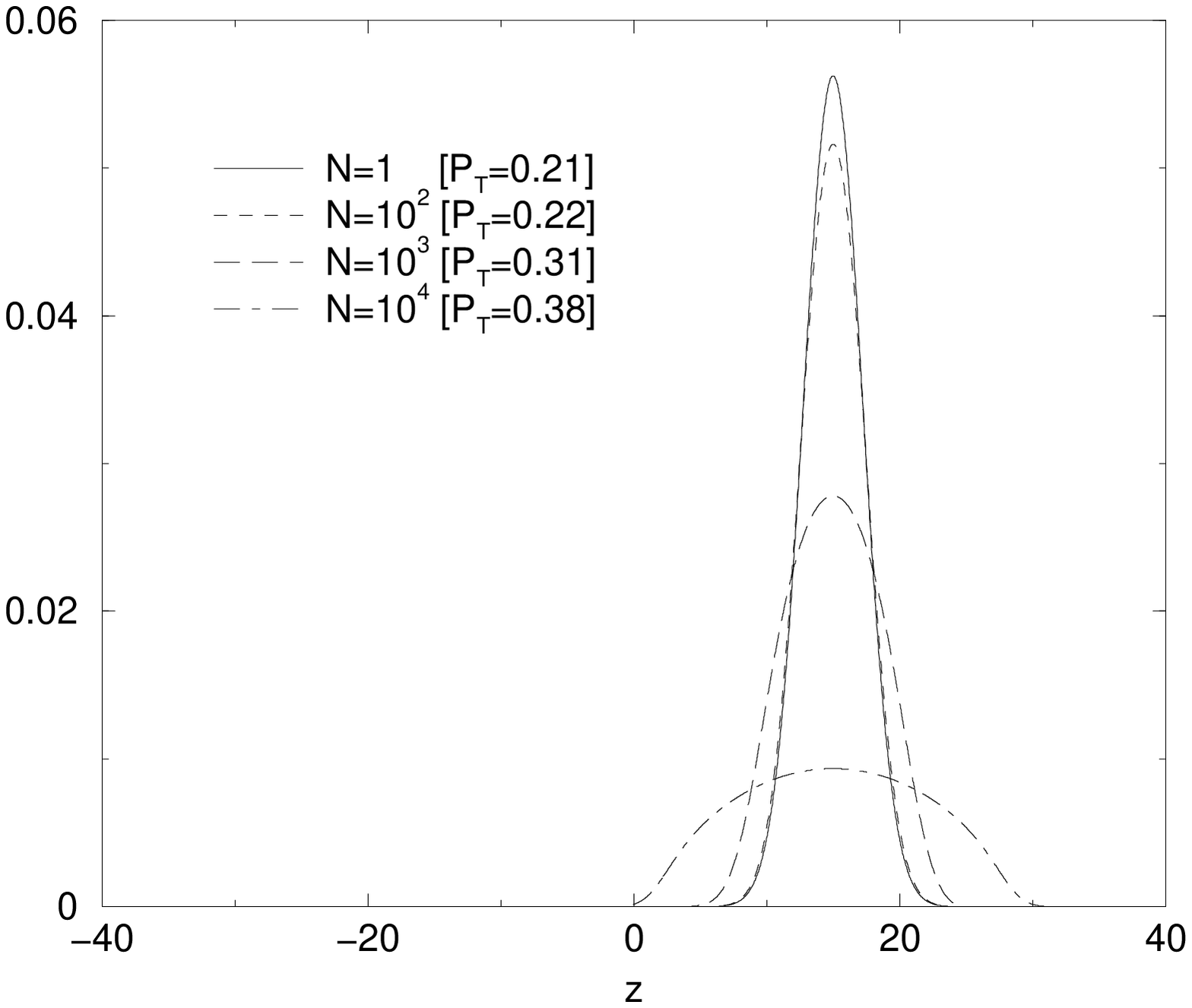,height=5.in}}
{\bf Fig. 4}: {Density profile $|\psi(\rho=0,z)|^2$ at $t=0$ 
for various values of the number $N$ of condensed atoms.  
Gaussian energy barrier $U e^{z^2/\sigma^2}$, 
with $U=200$ and $\sigma=1$. 
Radially-confined ($\omega_{\rho}=1$) falling condensate 
with cigar-shaped ($\omega_{\rho}/\omega_{z}=10$) 
initial condition. $P_T$ is the tunneling fraction. 
Units as in Fig 1.} 
\end{figure} 

\newpage

\begin{figure} 
\centerline{\psfig{file=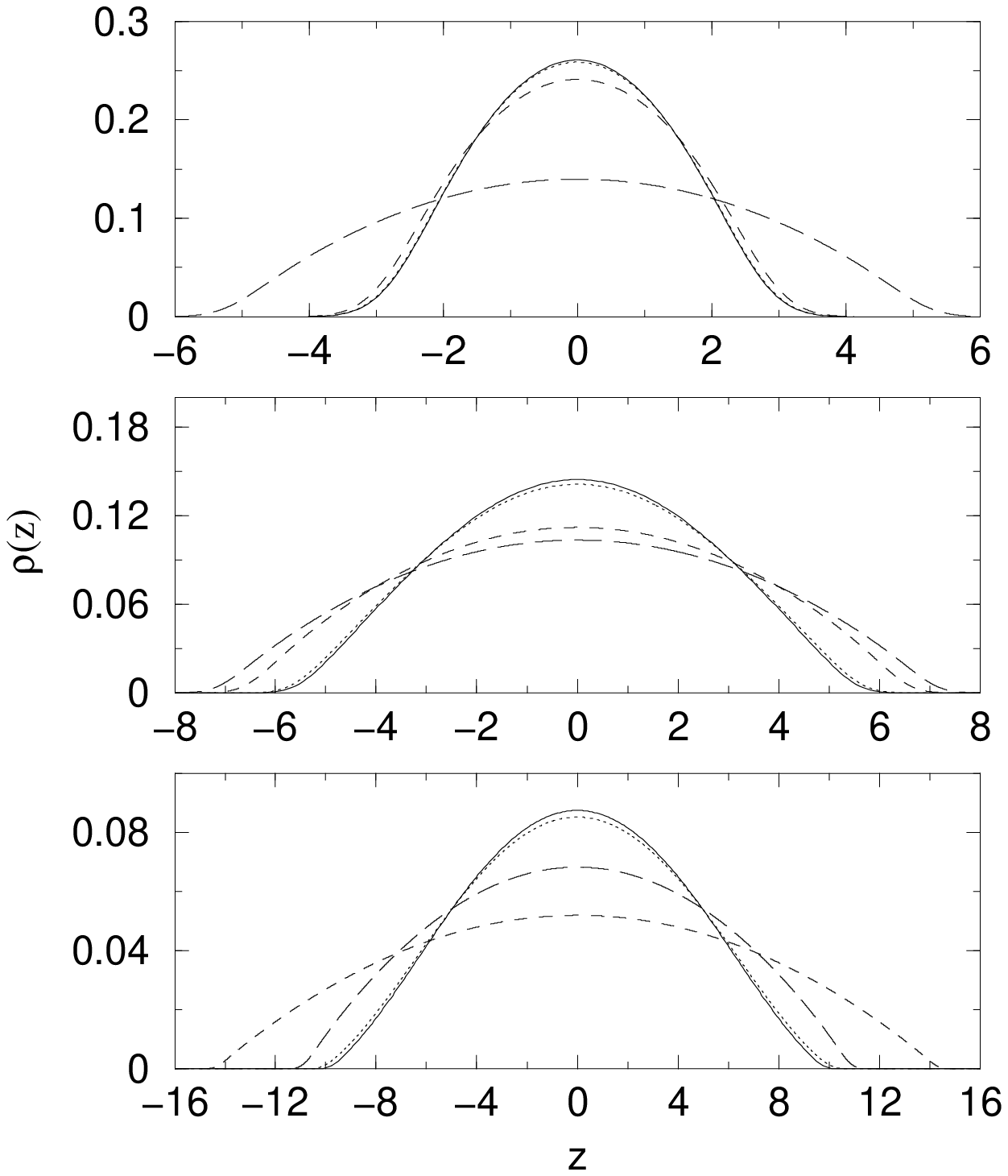,height=5.in}}
{\bf Fig. 5}: 
{Normalized density profile $\rho(z)=|f(z)|^2$ along the axial 
direction $z$ of the Bose condensate in harmonic potential. 
Number of Bosons: $N=10^4$ and 
trap anisotropy: $\omega_{\bot}/\omega_z=10$. 
Four different procedures: 
3D GPE (solid line), 1D GPE (dashed line), CGPE 
(long-dashed line) and NPSE (dotted line). 
From top to bottom: $a_s/a_z=10^{-4}$, 
$a_s/a_z=10^{-3}$, $a_s/a_z=10^{-2}$. 
Length $z$ in units of $a_z=\sqrt{\hbar/(m\omega_z)}$.} 
\end{figure} 

\end{document}